\documentclass[conference]{IEEEtran}
\IEEEoverridecommandlockouts

\usepackage{cite}
\usepackage{amsmath,amssymb,amsfonts}
\usepackage{algorithmic}
\usepackage{graphicx}
\usepackage{textcomp}
\usepackage{comment}
\usepackage{xcolor}
\usepackage{cuted}
\usepackage{capt-of}

\def\BibTeX{{\rm B\kern-.05em{\sc i\kern-.025em b}\kern-.08em
    T\kern-.1667em\lower.7ex\hbox{E}\kern-.125emX}}

\begin{document}

\title{
    LLM-Assisted Workflow for Structural Difference Visualization in Evolving Software Requirements
}

\makeatletter
\newcommand{\linebreakand}{%
  \end{@IEEEauthorhalign}
  \hfill\mbox{}\par
  \mbox{}\hfill
  \begin{@IEEEauthorhalign}
}
\makeatother

\author{
    \IEEEauthorblockN{Koi McFarland}
    \IEEEauthorblockA{
    \textit{Charleston Southern University, United States}\\
    kemcfarland@student.csuniv.edu}
    \and
    \IEEEauthorblockN{Songhui Yue}
    \IEEEauthorblockA{
    \textit{Charleston Southern University, United States}\\
    syue@csuniv.edu}
}

\maketitle

% ============================================================================================================
% Abstract
% ============================================================================================================
\begin{abstract}
This paper presents an LLM-assisted workflow for visualizing structural differences in evolving software requirements. Implemented in the OntologyWeb environment, the workflow represents baseline and current requirements as triple-based semantic graphs and supports side-by-side comparison of curated graph snapshots. The comparison view aligns matched entities and uses visual encoding to highlight structural changes.
\end{abstract}

\begin{comment}
keywords:
Structural Difference Visualization
Requirement Traceability
Change Impact Analysis
Large Language Models
Ontology
\end{comment}
% ============================================================================================================
% Introduction
% ============================================================================================================

\section{Introduction}
Software requirements often evolve as stakeholders refine goals, add constraints, or modify system behaviors. During review and maintenance, they may need to understand not only which words have changed, but also how the underlying semantics and relationships have changed. The capture and representation of these changes is related to long-standing work on requirements traceability \cite{cleland2014software} and change impact analysis \cite{li2008requirement}. Traditional textual comparisons expose mainly surface-level updates and may not clearly show structural changes such as added responsibilities, removed dependencies, or modified relationships among requirement concepts.  

Visualization has been studied as a way to help stakeholders inspect requirement artifacts and their relationships \cite{reddivari2014visual}. Ontology- and knowledge-graph-based representations provide a possible way to inspect such structural information. By representing requirement statements as triples, analysts can examine concepts and relations more explicitly than in plain text. Recent progress in Large Language Models (LLMs) has made it increasingly feasible to generate candidate triples and ontology structures from textual materials such as requirement documents \cite{yue2026llm}. However, this creates a workflow challenge: LLM-generated structures must be examined and validated by human experts, preserved as comparable versions, and visualized in a way that supports structural change analysis.

Existing ontology engineering and visualization tools, such as Protégé \cite{musen2015protege} and its graph-difference plugins, provide support for ontology editing, inspection, and comparison. However, they are not primarily organized around a requirement-version comparison workflow. Our workflow in the OntologyWeb tool is designed for this specific use case: comparing evolving requirements after they have been converted into triple-based or ontology graphs. The workflow combines LLM-assisted generation, human curation, snapshot selection, evidence inspection, and a side-by-side synchronized comparison layout that places matched entities in similar visual positions across baseline and current snapshots.

% ============================================================================================================
% Methodology
% ============================================================================================================

\begin{figure*}[htp]
    \centering
    \includegraphics[width=\textwidth]{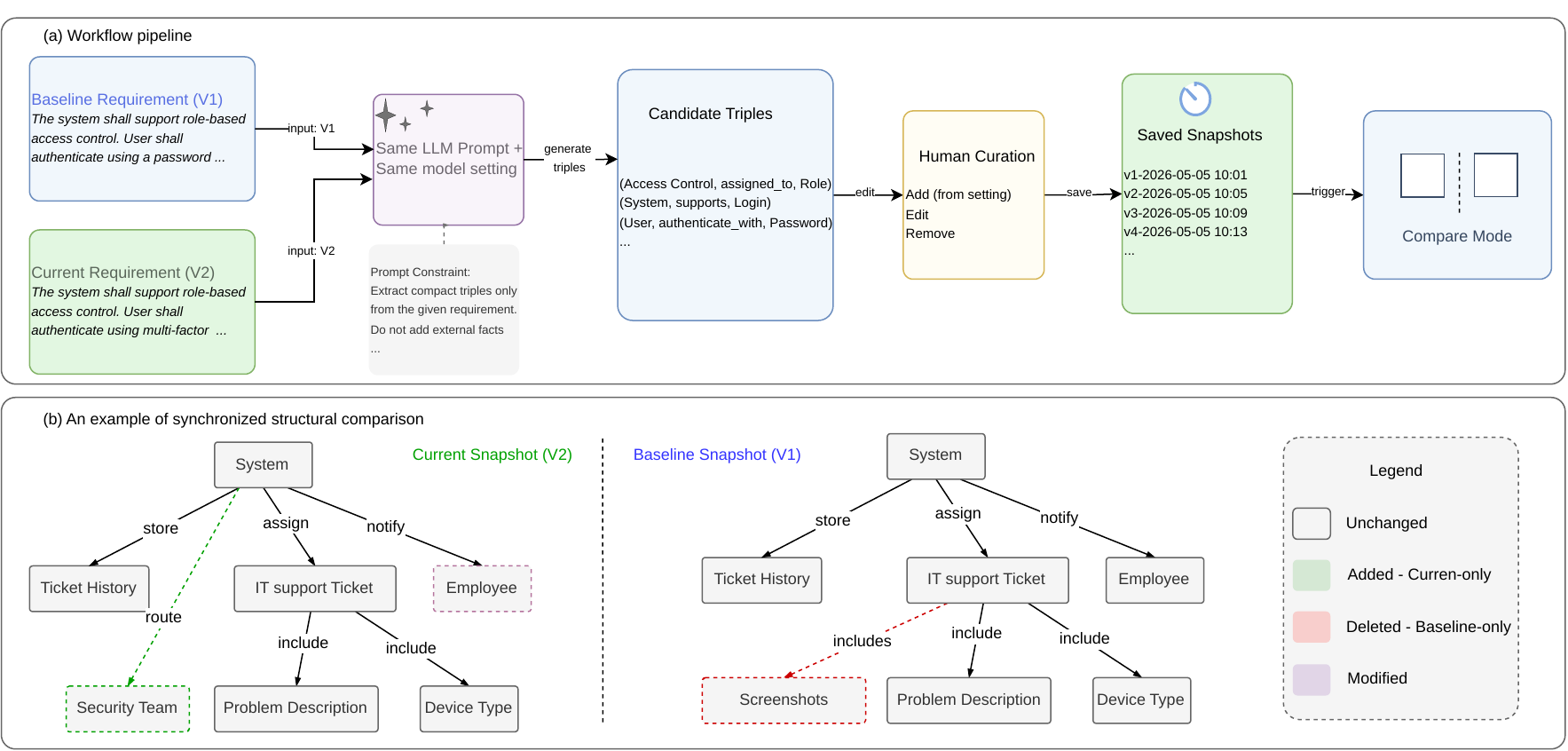}
    \caption{Overview of the proposed workflow: (a) steps for visualizing structural differences in evolving requirements and (b) a graph comparison example.}
    \label{fig:myfigure}
\end{figure*}

\section{Workflow for Visualizing Structural Differences}

\subsection{Workflow Overview}
The proposed workflow is implemented within the OntologyWeb environment, a web-based visualization-oriented tool that supports human-in-the-loop curation of LLM-generated triples and ontology fragments. In this paper, we focus on one requirements-engineering use case: using the tool to compare evolving requirement versions and visualize the structural differences.

The workflow connects four steps, as shown in Figure~\ref{fig:myfigure} (a). First, baseline and current requirement versions are processed using the same LLM prompt and model settings to generate candidate triples. Second, users inspect and curate the generated triples by editing, adding, or removing nodes and relations. Third, curated graph versions are saved as snapshots. Finally, the user selects two snapshots and compares them in a side-by-side graph view. This process supports structural difference analysis by helping users identify common, added, removed, and modified structures across requirement versions.

\subsection{Snapshot-based Comparison}
To support visualization of structural differences, the system can compare two curated graph snapshots generated from different versions of the same requirement. Figure~\ref{fig:myfigure} (b) shows an example comparison view in which shared concepts are aligned, and structural differences are highlighted.

The comparison view provides several interaction features: 1) The canvas uses a synchronized layout, where matched nodes appear in approximately similar positions across the baseline and current graphs. 2) Visual encoding, such as colored node or edge borders, is used to indicate structural changes such as addition, removal, and modification. 3) The user can also inspect evidence text by hovering over relations, allowing each generated relation to remain traceable to the original requirement text.

\subsection{Implementation and Demonstration}
The system is implemented as a web-based tool that integrates graph processing, interactive visualization, and an LLM pipeline. Graph structures are represented as subject-predicate-object triples and stored in a centralized JSON structure, which acts as the source for all transformations and rendering. To support matching, node and edge labels are normalized, and namespace prefixes are ignored when appropriate. We created a short 7-minute video demonstrating the tool on YouTube at this link: https://www.youtube.com/watch?v=Q3QH63-B8k0.

% ============================================================================================================
% Evaluation
% ============================================================================================================

\section{Preliminary Observations}
We conducted a preliminary qualitative assessment using real-world requirement documents, each with at least two versions. For each pair, the baseline and current versions were processed through the workflow to generate candidate triple-based graphs, save snapshots, and compare the resulting structures. The assessment was exploratory and focused on whether the visualization can help users inspect added, removed, and modified relations.

Our observations suggest that the workflow is more useful for medium- and high-complexity requirement changes than for short, simple changes. For short paragraphs, a direct textual comparison may be sufficient, while the graph-based workflow may require additional inspection effort. For complex requirements, the side-by-side visualization appears to make relations, dependencies, and changed structures more explicit. Future evaluation will use metrics such as task completion time, accuracy in identifying structural differences, user confidence in identifying the differences, and subjective usability ratings.

% ============================================================================================================
% Discussion and Conclusion
% ============================================================================================================
\section{Discussion and Conclusion}
The comparison workflow is currently in a prototype stage and focuses on graph-based presentation and highlighting of structural differences. A current limitation is that many requirements are stored as multi-paragraph documents, while the present workflow is more suitable for comparing selected short sections. One future direction is to support paragraph-by-paragraph comparison through additional workflow design.

The primary contribution of this work is not a new ontology-difference algorithm, but an integrated requirement-engineering-oriented workflow that connects LLM-assisted requirement structuring with interactive graph comparison. The workflow contributes: (1) a human-in-the-loop process for generating and curating requirement triples from different requirement versions; (2) a snapshot-based comparison mechanism that allows users to select and compare the generated graphs; (3) a synchronized side-by-side visualization that uses color encoding, evidence inspection, and layout alignment to help users inspect structural differences.

% ============================================================================================================
% Acknowledgment
% ============================================================================================================
\section*{Acknowledgment}
This material is based upon work supported by the NSF under Grant No. 2451397. AI tools, including ChatGPT, Gemini, and Grammarly, were used to support prototype refinement based on the authors' design specifications and to improve sentence conciseness and grammatical correctness.

\bibliographystyle{IEEEtran}
\bibliography{refs}

\end{document}